\documentclass[
twocolumn,
floatfix,superscriptaddress,a4paper,showkeys,nofootinbib,reprint]{revtex4-1}
\usepackage[colorlinks=true,linktocpage=true,linkcolor=blue,citecolor=blue,allcolors=blue]{hyperref}
\usepackage{epsfig}
\usepackage{latexsym}
\usepackage[utf8]{inputenc}
\usepackage{xspace}
\usepackage{indentfirst}
\usepackage{enumitem}
\usepackage{color}
\usepackage{placeins}
\usepackage[table,xcdraw]{xcolor}
\usepackage{setspace}
\usepackage{lipsum}
\usepackage{multirow}
\usepackage{hyperref}

\usepackage{graphicx}

\usepackage{todonotes}

\usepackage{amsmath}
\usepackage{amssymb}
\usepackage[english]{babel}
\usepackage{url}
\graphicspath{{figs/}}
\newcommand{\eq}[1]{\begin{align} #1 \end{align}}

\newcommand{\be}{\begin{equation}}
\newcommand{\ee}{\end{equation}}

\begin{document}

\title{Correlations between nuclear incompressibility, liquid-gas critical point, and quarkyonic transition}

\author{Artemiy Lysenko}
\affiliation{Physics Department, Taras Shevchenko National University of Kyiv, 03022 Kyiv, Ukraine}
\affiliation{Bogolyubov Institute for Theoretical Physics, 03680 Kyiv, Ukraine}

\author{Mark I. Gorenstein}
\affiliation{Bogolyubov Institute for Theoretical Physics, 03680 Kyiv, Ukraine}

\author{Tripp~Moss}
\affiliation{Physics Department, University of Houston, Box 351550, Houston, TX 77204, USA}

\author{Roman Poberezhniuk}
\affiliation{Physics Department, University of Houston, Box 351550, Houston, TX 77204, USA}
\affiliation{Bogolyubov Institute for Theoretical Physics, 03680 Kyiv, Ukraine}

\author{Volodymyr Vovchenko}
\affiliation{Physics Department, University of Houston, Box 351550, Houston, Texas 77204, USA}

\begin{abstract}
 We systematically probe different parametrizations of the attractive nuclear force based on real gas models to construct the nuclear matter equation of state. In each of the cases, the repulsion between nucleons is treated in the framework of excluded volume, and interaction parameters are fitted to the empirical properties of the nuclear ground state. We calculate the critical temperature $T_c$ and critical particle number density $n_c$, and find that they are strongly correlated. Both are also correlated with the incompressibility $K_0$ in the nuclear ground state. We also include a quarkyonic matter phase in the quasiparticle description and investigate the relationships among $K_0$, transition density to the quarkyonic phase, $n_{tr}$, and corresponding peak in the speed of sound, $v_{s, {\rm max}}^2$. At each density, the quark fraction is found by minimizing the energy density. We find that both $n_{tr}$ and $v_{s, {\rm max}}^2$ are negatively correlated with $K_0$, $n_c$, and $T_c$.
\end{abstract}

\maketitle

\section{Introduction}

The study of the nuclear matter equation of state (EoS) is fundamental to understanding the properties of strongly interacting matter across a wide range of densities and temperatures. At subnuclear densities, nuclear matter is well described by effective models incorporating the interplay of attractive and repulsive interactions between nucleons~\cite{Negele:1971vb,Serot:1984ey}. These interactions are critical in reproducing empirical properties of the nuclear ground state, such as the saturation density $n_0$, binding energy $\varepsilon/n_0-m$, and incompressibility $K_0$. At higher densities, where nucleons begin to overlap, the emergence of quark degrees of freedom leads to the transition to new phases of matter, such as quarkyonic matter~\cite{McLerran:2007qj,McLerran:2018hbz,Philipsen:2019qqm,Kovensky:2020xif}.

The van der Waals (vdW) EoS \cite{Landau:1980mil} was introduced in 1873 for real gases. The two parameters $a$ and $b$ in this equation represent attractive and repulsive interactions, respectively. The vdW EoS was first used for a description of nuclear matter in Ref.~\cite{Vovchenko:2015vxa}, where the effect of Fermi statistics was introduced. The model predicts a first-order liquid-gas phase transition and a nuclear critical point (CP) with critical temperature $T_c$ and density $n_c$ in reasonable agreement with empirical estimates. 

Although the vdW EoS overestimates ground state incompressibility, $K_0$, the agreement with experimental data can be improved by using modifications of the vdW EoS based on other real gas models~\cite{Vovchenko:2017cbu}. In this work, we systematically explore a variety of parametrizations of the attractive nuclear force using real gas models from~\cite{Vovchenko:2017cbu}, including Redlich-Kwong-Soave, Peng-Robinson, and Clausius models, and explore a new attraction parametrization based on the Dieterici EoS for real gases~\cite{dieterici1899ueber}. In the case of two-parameter Clausius and Dieterici attraction terms, we vary the parameters such that they interpolate smoothly between vdW EoS and the original real gas models. For each of the considered equation of state the strength of both repulsion and attraction are uniquely chosen to satisfy the empirical ground state properties of nuclear matter.

Of course, there are more first-principle-based approaches such as chiral effective field theory~\cite{Drischler:2021kxf} that are arguably more appropriate for the description of nuclear matter. The class of real gas model that we explore has a couple of advantages; however, (i) it can be extended in a relatively straightforward manner to multicomponent systems such as hadron resonance gas and (ii) one can introduce a transition to quarkyonic matter regime. These extensions have already been done for the case of vdW interaction in Refs.~\cite{Vovchenko:2016rkn, Vovchenko:2017zpj} and~\cite{Poberezhnyuk:2023rct,Moss:2024uam}, respectively.

The analysis of these different models shows a distinctive correlation between the $K_0$ parameter and $T_c$ values: the stiffer nuclear matter, i.e., larger $K_0$, the larger the values of $T_c$ and $n_c$~\cite{Kapusta:1984ij,Lattimer:1991nc,Natowitz:2002nw,Rios:2010jh}. 
For instance, this correlation was investigated in Ref.~\cite{Lourenco:2016jdo} within a relativistic mean-field model. 
In the present paper we study an interplay of the repulsive and attractive interactions between nucleons responsible for the observed correlations. We also emphasize the effects of quantum statistics. The Fermi statistics of nucleons is not only crucial at zero temperature, but is also important in the vicinity of the CP.

We extend our analysis to higher baryon densities by including quark degrees of freedom in a quasiparticle picture for quarkyonic matter~\cite{McLerran:2018hbz,Jeong:2019lhv}. In this picture at $T=0$ quarks fill the Fermi sea while baryonic excitations exist around the Fermi surface. The resulting EoS models the effects of Pauli exclusion principle between constituent quarks which becomes relevant at higher densities. The transition to the quarkyonic regime is characterized by a peak in the speed of sound $v_{s,\rm{max}}^2$ at a density $n_{tr}$ where the transition occurs. This peak exceeds the conformal limit, which is supported by neutron star observations~\cite{Tews:2018kmu,Fujimoto:2019hxv,Tan:2020ics,Altiparmak:2022bke}. Here we extend the vdW model analysis of~\cite{Poberezhnyuk:2023rct,Moss:2024uam} to real gas models and observe anti-correlation between $K_0$ and both $n_{tr}$ and $v_{s,\rm{max}}^2$.

The paper is organized as follows. Five different models for real gases are presented in Sec.~\ref{eos}. The correlations between $K_0$, $T_c$ and $n_c$ are described in Sec.~\ref{cor}. The correlation between $K_0$ and the onset density of quarkyonic matter is discussed in Sec.~\ref{quarkyonic_matter}. A summary in Sec.~\ref{sum} closes the article.

\section{Equations of state of real gases}\label{eos}

Classical equations of state for real gases are usually presented in terms of the pressure $P$ as a function  of temperature $T$ and particle number density $n\equiv N/V$. In Ref.~\cite{Vovchenko:2017cbu}\footnote{References to original papers that introduce equations of state (\ref{P_VDW})--(\ref{P_CTn}) are presented in Ref. \cite{Vovchenko:2017cbu}.} several examples of real gas models were used to calculate the properties of nuclear matter:
\begin{itemize}
\item van der Waals (vdW) 
\eq{ \label{P_VDW}
P_{\rm VDW}(T,n) = \frac{nT}{1 - bn} - an^2~,
}
    \item Redlich-Kwong-Soave (RKS) 
    \eq{ \label{P_RKSTn}
       P_{\rm RKS}(T, n) = \frac{nT}{1 - bn} - \frac{an^2}{1 + bn}~,
    }
    \item Peng-Robinson (PR) 
    \eq{ \label{P_PRTn}
       P_{\rm PR}(T, n) = \frac{nT}{1 - bn} - \frac{an^2}{1 +2bn- (bn)^2}~, 
       }
       \item and Clausius 
    \eq{ \label{P_CTn}
       P_{\rm C}(T, n) = \frac{nT}{1 - bn} - \frac{an^2}{(1 + cn)^2} ~.
       }
\end{itemize}
The first term in Eqs.~(\ref{P_VDW})--(\ref{P_CTn}) include repulsive interactions in the framework of excluded volume. Here the parameter $b$ regulates the strength of these interactions. The second term in Eqs.~(\ref{P_VDW})--(\ref{P_CTn}) describes the different forms of mean-field attraction regulated by the parameter $a$. Note that Eq.~(\ref{P_CTn}) includes also a third parameter $c\ge 0$, which additionally regulates attractive interactions, suppressing them with increasing $c$. At $c=0$ the Clausius model is reduced to the vdW model.  

In addition to the above equations for real gases we consider the Dieterici EoS. For our purposes we formulate the generalized Dieterici equation in the following form ($5/3\leq \alpha \leq 2$):
\eq{ \label{P_DTn}
   P_{\rm D}(T, n) = \frac{nT}{1 - bn} - an^{\alpha}~.
}
Dieterici originally proposed this equation in 1899~\cite{dieterici1899ueber} with $\alpha=5/3$. Here we use Dieterici equation in a generalized form with arbitrary $\alpha$ parameter which allows a smooth transition from the original Dieterici model ($\alpha=5/3$) to the vdW  model ($\alpha=2$).\footnote{We note that there exists another common variant of the Dieterici equation of state with an exponential term~\cite{10.1063/1.1380711}, which is distinct from the one studied here.}
  
While repulsion in all listed models is described identically with the excluded volume of vdW, the attractive forces have different functional dependence on particle number density $n$.
The parameters $a$ and $b$ can, in principle, be related to the lowest order coefficient of the virial expansion~\cite{Landau:1980mil,Vovchenko:2017drx}, which is constrained by empirically known scattering phase shifts (see, e.g., Refs.~\cite{Wiringa:1994wb, Horowitz:2005zv}). However, in the baryon-rich matter considered here, multibody effects are dominant near the liquid-gas transition that we study here. Therefore, the interaction parameters are not constrained by the two-body physics of the virial expansion but instead fitted to empirical properties of the nuclear ground state, providing an effective description of the multibody effects relevant there.

\subsection{Quantum statistics and GCE}

The real gas equations (\ref{P_VDW})--(\ref{P_DTn}) correspond to classical particles.\footnote{The CP $T_c$ and $n_c$ values can be calculated analytically in this case, and they are presented in the Appendix.} Such a description fails at low temperatures and/or high particle densities. Thus, the first step to use them for nuclear matter is to introduce Fermi statistics for nucleons. This procedure can be most conveniently done in the grand canonical ensemble (GCE), and it is described in detail in Ref.~\cite{Vovchenko:2017cbu}. We present below only the final formulas. The GCE pressure for Eqs.~(\ref{P_VDW})--(\ref{P_DTn}) is given by the following system of transcendental equations:
\eq{ \label{P-Tmu}
   P(T, \mu) & = p^{\rm id}(T, \mu^*) + n^2u'(n)~, \\
   \label{e-Tmu}
   \varepsilon(T, \mu) & = (1-bn)\varepsilon^{\rm id}(T, \mu^*) + n u(n) \\
  \label{n-Tmu}
   n(T, \mu) & = (1-bn)n^{\rm id}(T, \mu^*), \\ 
   \label{mu*}
   \mu^{*} & = \mu - b p^{\rm id}(T, \mu^{*}) - u(n) - nu'(n)~,
} 
where $\mu$ is the chemical potential, $\mu^*$ is the effective chemical potential, $\varepsilon$ is energy density, and the function $u(n)$ corresponds to the energy of attractive interactions per particle.  For the above models $u(n)$ is given as
\eq{\label{u-n}
& u_{\rm VDW}(n)=-an~,~~~ u_{\rm RKS}(n)=- \frac{a}{b}~\ln(1+bn)~,~\\
& u_{\rm PR}(n)= -~\frac{a}{2\sqrt{2}b}~\ln\frac{1+bn+\sqrt{2}bn}{1+bn-\sqrt{2}bn}~,
\label{u-n-1}\\
&u_{\rm C}(n)=-~\frac{an}{1+cn}~,~~~  u_{\rm D}(n)= -~a\frac{n^{\alpha - 1}}{\alpha - 1}~.
\label{u-n-2}
} 
In Eqs.~(\ref{P-Tmu})--(\ref{mu*}) the ideal Fermi gas  functions are calculated as \cite{Landau:1980mil} 
\eq{ \label{p-id}
   & p^{\rm id}(T, \mu) = \frac{d}{6\pi^{2}} \int_{0}^{\infty} dk \frac{k^{4}}{\sqrt{m^{2} + k^{2}}} \times \nonumber \\ & \times \left[\exp\left(\frac{\sqrt{m^{2} + k^{2}} - \mu}{T}\right) ~+~ 1 \right]^{-1}~, \\
   \label{n-Tmu}
 &  n^{\rm id}(T, \mu) = \left( \frac{\partial p^{\rm id}}{\partial \mu} \right)_T~, \\ 
 &  \varepsilon^{\rm id}(T, \mu) = T\left( \frac{\partial p^{\rm id}}{\partial T} \right)_\mu +\mu n^{\rm id} - p^{\rm id} ~, \label{e-Tmu-1}
 }  
 where $m\cong 938$~MeV is the nucleon mass and $d=4$ is the nucleon spin-isospin degeneracy factor.
  
\subsection{Ground state of nuclear matter}

\begin{table}[h!]
\begin{tabular}{| c | c | c | c | c | c |}
  \hline
  Model & $a \ \rm MeV \ fm^{3}$ & $b \ \rm fm^{3}$ & $K_0 \ \rm MeV$ & $T_c \ \rm MeV$ & $n_c \ \rm fm^{-3}$ \\
  \hline
  VDW & $329$ & $3.41$ &  $763$ & $19.7$ & $0.072$ \\
  RKS & $374$ & $2.94$ &  $518$ & $18.2$ & $0.064$ \\
  PR & $408$ & $2.82$ &  $443$ & $17.1$ & $0.061$ \\
  \hline 
\end{tabular}
\caption{The values of the VDW-like parameters $a$ and $b$, the values of the resulting nuclear incompressibility, and the properties of the CP of nuclear matter for vdW, RKS, and PR models.}
\label{table-1}
\end{table}

\begin{table*}
\begin{tabular}{| c | c | c | c | c | c | c | c |}
  \hline
  $\alpha$ & $an_0^{\alpha} \ \rm MeV \ fm^{-3}$ & $b \ \rm fm^{3}$ & $K_0 \ \rm MeV$ & $T_c \ \rm MeV$ & $n_c \ \rm fm^{-3}$ & $v^2_{s,{\rm max}}$ & $n_{tr}$ (units of $n_0$) \\
  \hline
  $5/3$ & $4.86$ & $2.28$ &  $250$ & $13.0$ & $0.058$ & $0.686$ & $2.13$ \\
  $1.70$ & $5.19$ & $2.43$ &  $288$ & $13.7$ & $0.060$ & $0.665$ & $2.00$ \\
  $1.75$ & $5.67$ & $2.64$ &  $350$ & $14.8$ & $0.063$ & $0.639$ & $1.84$ \\
  $1.80$ & $6.20$ & $2.83$ &  $418$ & $15.8$ & $0.065$ & $0.622$ & $1.72$ \\
  $1.85$ & $6.74$ & $3.00$ &  $494$ & $16.8$ & $0.067$ & $0.609$ & $1.62$ \\
  $1.90$ & $7.29$ & $3.15$ &  $576$ & $17.8$ & $0.069$ & $0.602$ & $1.54$ \\
  $1.95$ & $7.86$ & $3.29$ &  $666$ & $18.8$ & $0.071$ & $0.597$ & $1.48$ \\
  $2.00$ & $8.42$ & $3.41$ &  $763$ & $19.7$ & $0.072$ & $0.596$ & $1.43$ \\
  \hline
\end{tabular}
\caption{The values of interaction parameters $a$ and $b$, the resulting incompressibility $K_0$, critical temperature $T_c$, critical density $n_c$, speed of sound at the peak $v^2_{s,{\rm max}}$, and the quarkyonic transition density $n_{tr}$ in Dieterici model with different values of parameter $\alpha$.}
\label{table-2-1}
\end{table*}

\begin{table*}
\begin{tabular}{| c | c | c | c | c | c | c | c |}
  \hline
  $c \ \rm fm^3$ & \ \ \ $a \ \rm MeV \ fm^{3}$ \ \ & $b \ \rm fm^{3}$ & $K_0 \ \rm MeV$ & $T_c \ \rm MeV$ & $n_c \ \rm fm^{-3}$ & $v^2_{s,{\rm max}}$ & $n_{tr}$ (units of $n_0$) \\
  \hline
  $4.74$ & $472$ & $1.73$ &  $250$ & $16.2$ & $0.050$ & $0.742$ & $2.79$ \\
  $4.00$ & $451$ & $1.98$ &  $286$ & $16.5$ & $0.052$ & $0.692$ & $2.44$ \\
  $3.00$ & $422$ & $2.32$ &  $350$ & $17.1$ & $0.056$ & $0.645$ & $2.08$ \\
  $2.00$ & $392$ & $2.67$ &  $438$ & $17.8$ & $0.061$ & $0.615$ & $1.81$ \\
  $1.50$ & $376$ & $2.85$ &  $496$ & $18.2$ & $0.063$ & $0.604$ & $1.70$ \\
  $1.00$ & $361$ & $3.04$ &  $566$ & $18.7$ & $0.066$ & $0.597$ & $1.59$ \\
  $0.50$ & $345$ & $3.22$ &  $653$ & $19.2$ & $0.069$ & $0.595$ & $1.51$ \\
  $0.00$ & $329$ & $3.41$ &  $763$ & $19.7$ & $0.072$ & $0.596$ & $1.43$ \\
  \hline
\end{tabular}
\caption{The same as in Table \ref{table-2-1} but for Clausius model with different values of parameter $c$.}
\label{table-2-2}
\end{table*}

The model parameters are constrained by nuclear ground state properties. In the nuclear ground state $T=0$  and the total energy has a minimum (correspondingly $P=0$) with the following values of nucleon number density $n_0$ and binding energy per nucleon $W_0$~\cite{annurev:/content/journals/10.1146/annurev.ns.21.120171.000521}:  
\eq{\label{n0}
& n_0 ~= ~0.16 ~ {\rm fm}^{-3}~,~\\
& W_0 ~\equiv ~ \frac{\varepsilon (T=0,n=n_0)}{n_0}~-~m~=~-~16~{\rm MeV}~. \label{W0} 
}
The vdW, RKS, and PR models contain only two parameters, $a$ and $b$, and the ground state values of $n_0$ and $W_0$ fix them uniquely. Correspondingly, the CP location on the phase diagram and the value of incompressibility, $K_0 = 9(\partial P / \partial n)_{T=0}$,  become model predictions. The same is true for Dieterici and Clausius models at specified values of parameters $\alpha$ and $c$, respectively.

\begin{figure}[h!]
    \includegraphics[scale=0.4]{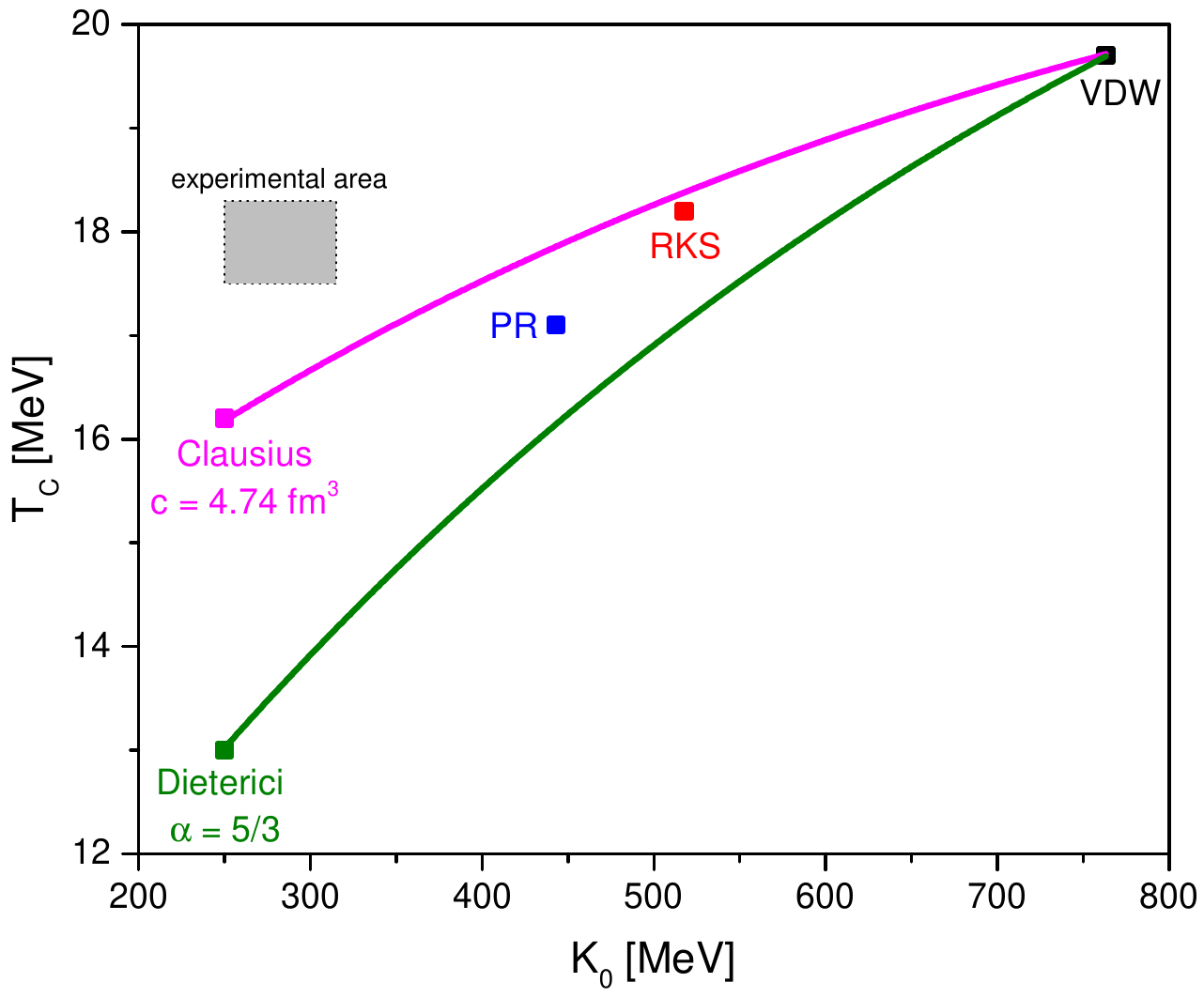}
    \caption{ 
    Critical temperature $T_c$ as a function of incompressibility $K_0$ for the VDW, RKS, PR, Dieterici, and Clausius models. The experimental estimates~\cite{Stone:2014wza, PhysRevC.87.054622} are shown by a gray rectangle.
    }
    \label{fig2}
\end{figure}

\section{Correlations among $K_0$, $T_c$, and $n_c$}\label{cor}

\begin{figure*}
    \includegraphics[width=.49\textwidth]{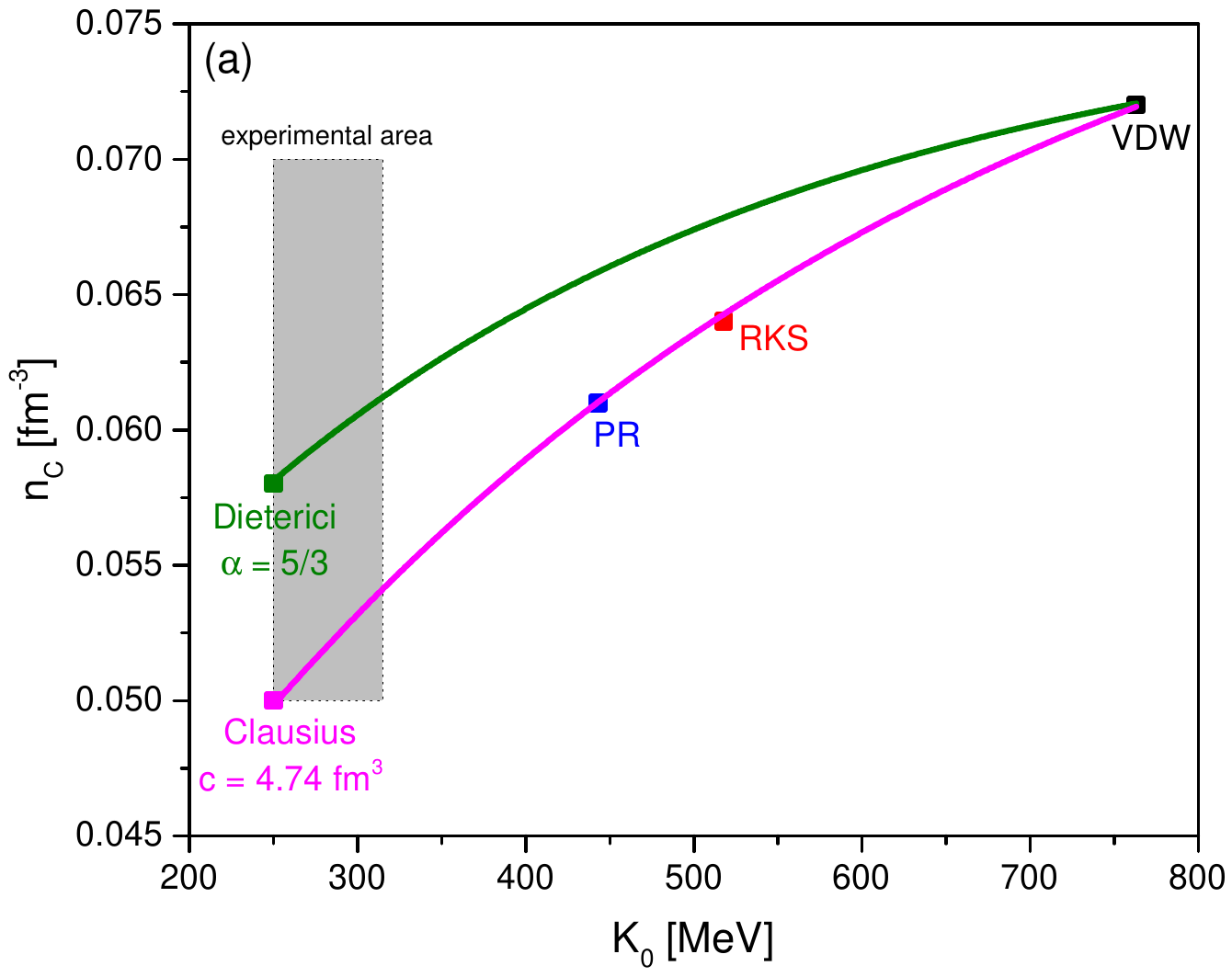}
    \includegraphics[width=.49\textwidth]{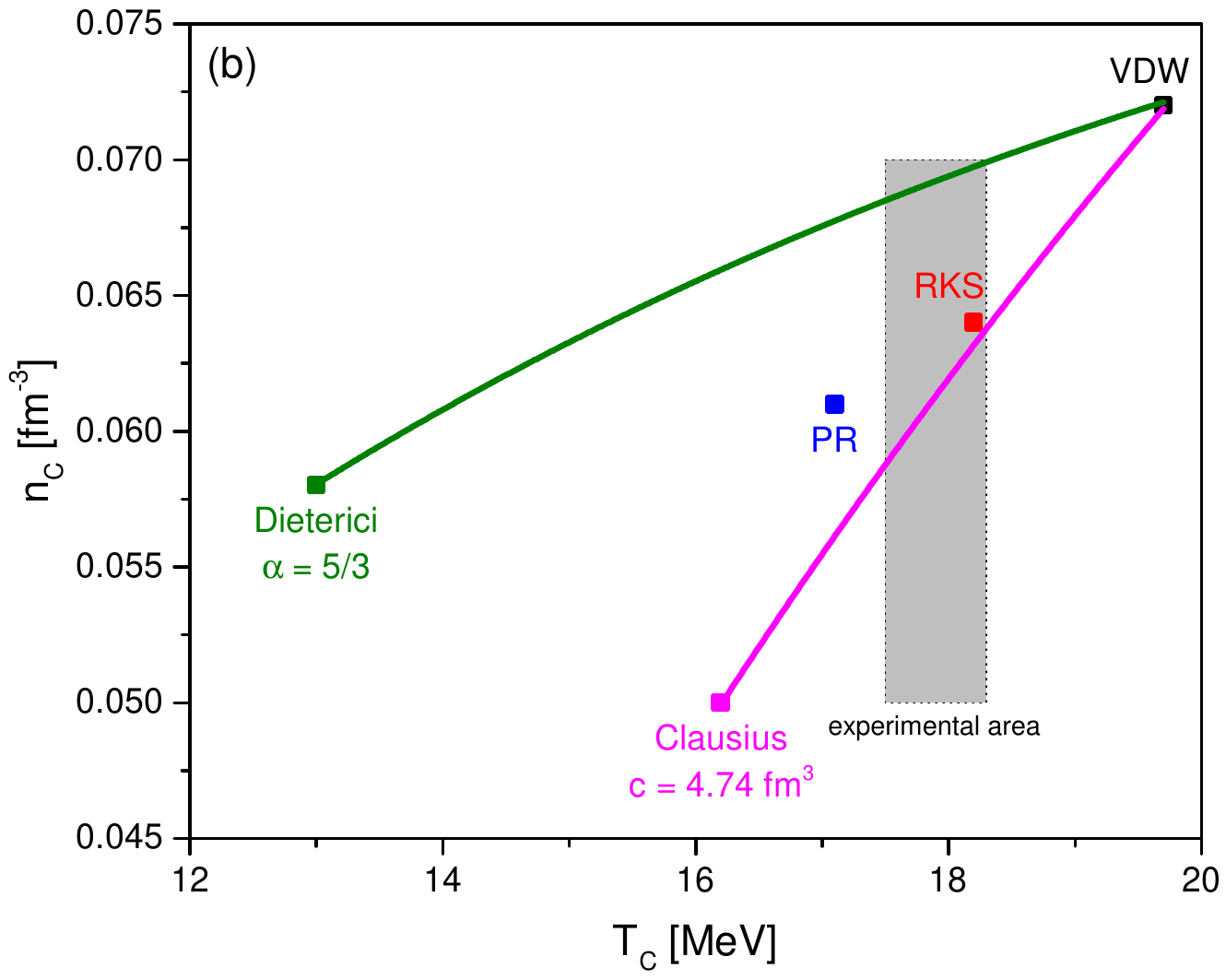}
    \caption{ 
     Critical density $n_c$ as the function of (a) incompressibility $K_0$ and (b) of the critical temperature $T_c$ in the vdW, RKS, PR, Clausius, and Dieterici models. The experimental estimates~\cite{Stone:2014wza, PhysRevC.87.054622} are shown by gray rectangles.
    }
    \label{fig3}
\end{figure*}

The location of the CP, $T_c$ and $n_c$, is defined by the following set of equations \cite{Landau:1980mil}:
\eq{\label{CP1}
\left( \frac{\partial P}{\partial n}\right)_T=0~,~~~~
\left( \frac{\partial^2 P}{\partial n^2}\right)_T=0 \ .
}
We treat $\alpha$ and $c$ as a free parameters within ranges $\alpha \in [5/3; \ 2]$ and $c \in [0; \ 4.74] \ \rm fm^3$. A complete set of model parameters fitted to Eqs.~(\ref{n0}) and (\ref{W0}) together with calculated $K_0$, $T_c$, and $n_c$ values for the vdW, RKS, and PR models are presented in Table~\ref{table-1}, and for Dieterici and Clausius models in Tables~\ref{table-2-1} and~\ref{table-2-2}, respectively. The corresponding experimental estimates are the following~\cite{Stone:2014wza, PhysRevC.87.054622}: $K_0 \in [250; 315] \ \rm MeV$, $T_c = 17.9\pm0.4 \ \rm MeV$, and $n_c = 0.06\pm 0.01 \ \rm fm^{-3}$.
These estimates are shown by gray rectangles in Figs.~\ref{fig2} and \ref{fig3}. We also note that neutron star observations and transport model calculations generally suggest a soft nuclear matter EoS with $K_0 < 300$~MeV (see Refs.~\cite{Wang:2018hsw,Huth:2020ozf,Sorensen:2023zkk} for details).

One sees that, while Clausius and Dieterici models are consistent with experimental data for $K_0$ and $n_c$, none of the considered models describe well $K_0$ and $T_c$ simultaneously. This issue can be mitigated by considering density-dependent excluded volume~\cite{Vovchenko:2017cbu} or mean fields~\cite{Lourenco:2019ist}, or attraction based on relativistic mean field~\cite{Poberezhnyuk:2017yhx}. However, here we focus on studying general correlations and leave these improvements for future work. 

The values of $T_c$ as a function of $K_0$ for all five considered models are shown in Fig.~\ref{fig2}. The green and pink lines represent, respectively the results for Dieterici and Clausius models with continuously varying free parameters $\alpha \in [5/3; \ 2]$ and $c \in [0; \ 4.74] \ \rm fm^3$. As expected, when $\alpha\rightarrow 2$ and $c\rightarrow 0$ the results for both models reduce to the vdW result.

The correlation between the critical temperature $T_c$ and the incompressibility $K_0$, shown in Fig.~\ref{fig2}, is well approximated by the formula $T_c = A\sqrt{K_0} + B$, with the $K_0^{1/2}$ dependence noted long ago~\cite{Kapusta:1984ij}. 
In the case of the Clausius model (\ref{P_CTn}), the approximation parameters $A$ and $B$ are as follows: $A = (0.302 \pm 0.005) \ \rm MeV^{1/2}$, $B = (11.449 \pm 0.105) \ \rm MeV$, and in the case of the Dieterici model (\ref{P_DTn}) $A = (0.570 \pm 0.007) \ \rm MeV^{1/2}$, $B = (4.082 \pm 0.157) \ \rm MeV$.

One observes a correlation  between $K_0$ and $T_c$, i.e., a larger $K_0$ corresponds to a larger $T_c$. A similar correlation exists between $K_0$ and the critical particle number density $n_c$; see Fig. \ref{fig3}(a). Correspondingly, $T_c$ and $n_c$ are also correlated; see Fig.~\ref{fig3}(b). These correlations were earlier studied in Refs.~\cite{Santos:2014vda, Santos:2015lma, Lourenco:2016jdo} within relativistic mean-field models.
 
Here, for all models the repulsive interactions are described by the excluded volume correction. This ensures that larger values of the parameter $b$ lead to to stronger repulsive effects. On the other hand, for all models the same ground state conditions (\ref{n0}) and (\ref{W0}) should be satisfied. Thus, the increase in repulsive interactions, i.e., the increase of $b$, has to be compensated by a corresponding increase of attractive forces. From Tables~\ref{table-1}--\ref{table-2-2} and Figs.~\ref{fig2} and \ref{fig3} one sees that, in all considered models, all three quantities, $T_c$, $n_c$, and $K_0$, increase when both the attractive and repulsive interactions increase consistently. 

\section{Correlation between nuclear incompressibility and the onset density of quarkyonic matter} \label{quarkyonic_matter}

The equations of state of real gases considered here can be employed for hadronic interactions in the model of quarkyonic matter~\cite{McLerran:2007qj,McLerran:2018hbz}. These interactions are essential for the dynamical generation of a momentum shell structure, as a noninteracting hadron gas is always energetically favorable to a free gas of quarks~\cite{Jeong:2019lhv}. Earlier, some of us studied the quantum van der Waals quarkyonic matter for isospin symmetric~\cite{Poberezhnyuk:2023rct} and asymmetric~\cite{Moss:2024uam} nuclear matter. There, only the vdW form~\eqref{u-n} of the mean-field attraction was studied with variations on the excluded volume prescription, leading to very stiff nuclear matter EoS with incompressibility $K_0 \sim 564-763$~MeV. The quarkyonic onset density in these models was predicted to be in the range $n \sim 1.5-2.0 n_0$ for symmetric matter.

Here, we utilize different prescriptions for attraction as discussed in Sec.~\ref{eos} with van der Waals excluded volume as the nucleonic part of the quarkyonic matter equation of state. This allows us to cover a broad range of nuclear incompressibilities, $K_0 \sim 250-763$~MeV, and study the correlation between $K_0$ and the quarkyonic matter onset density. We implement the quasiparticle picture which describes quarkyonic matter as a mixture of nucleons and quarks with Pauli exclusion principle acting between them. Correspondingly, the baryon density and energy density of quarkyonic matter are given as sums of nucleon and quark contributions, 
\eq{
n &= n_Q + n_N \ , \\
\varepsilon &= \varepsilon_Q + \varepsilon_N \ .
}

\begin{figure}[h!]
    \includegraphics[scale=0.4]{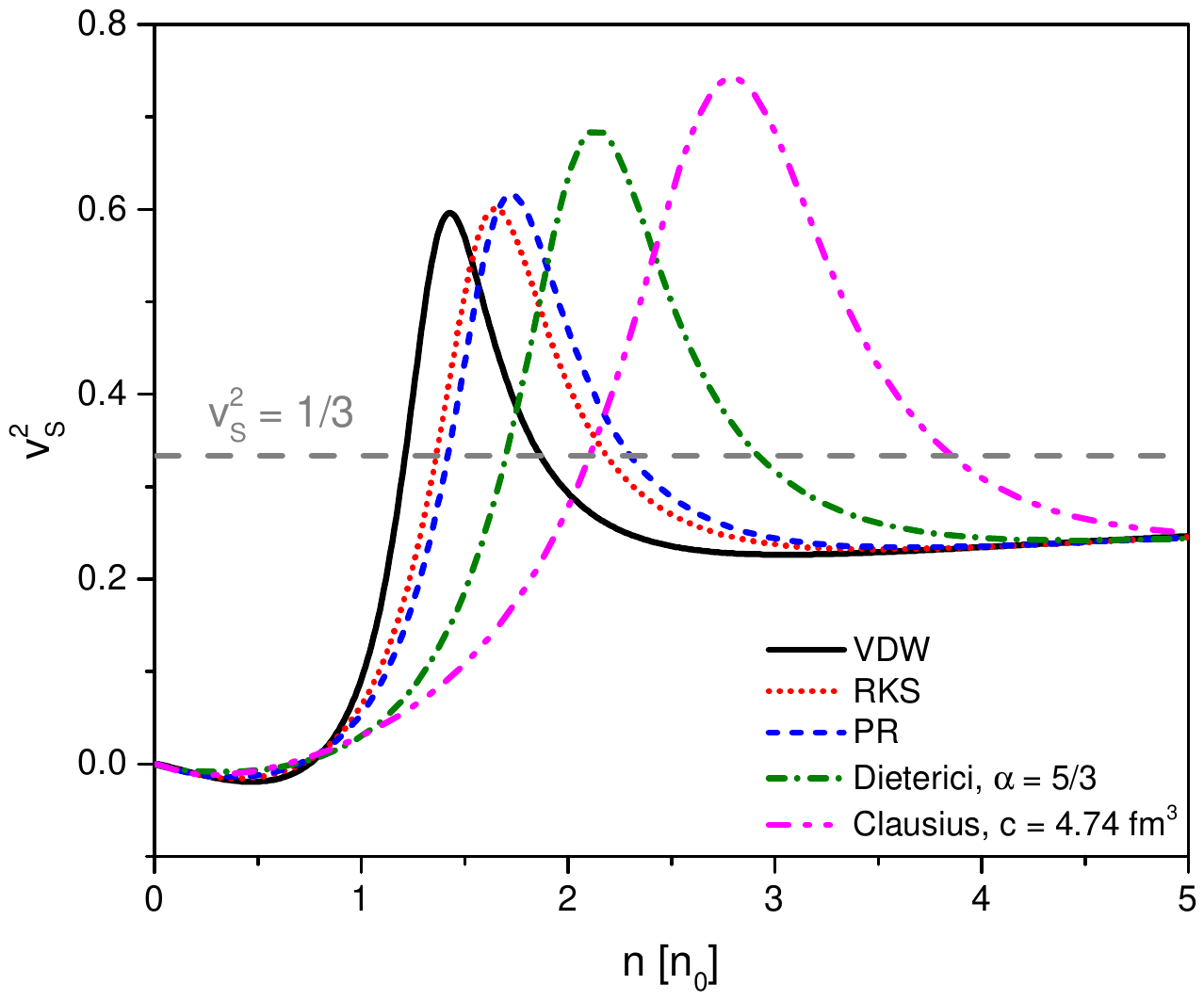}
    \caption{Speed of sound $v_s^2$ as a function of density $n$ in the vdW, RKS, PR, Clausius, and Dieterici models. The gray dashed line $v_s^2 = 1/3$ denotes the conformal limit.}
    \label{figv2snn0}
\end{figure}

\begin{figure*}
    \includegraphics[width=.49\textwidth]{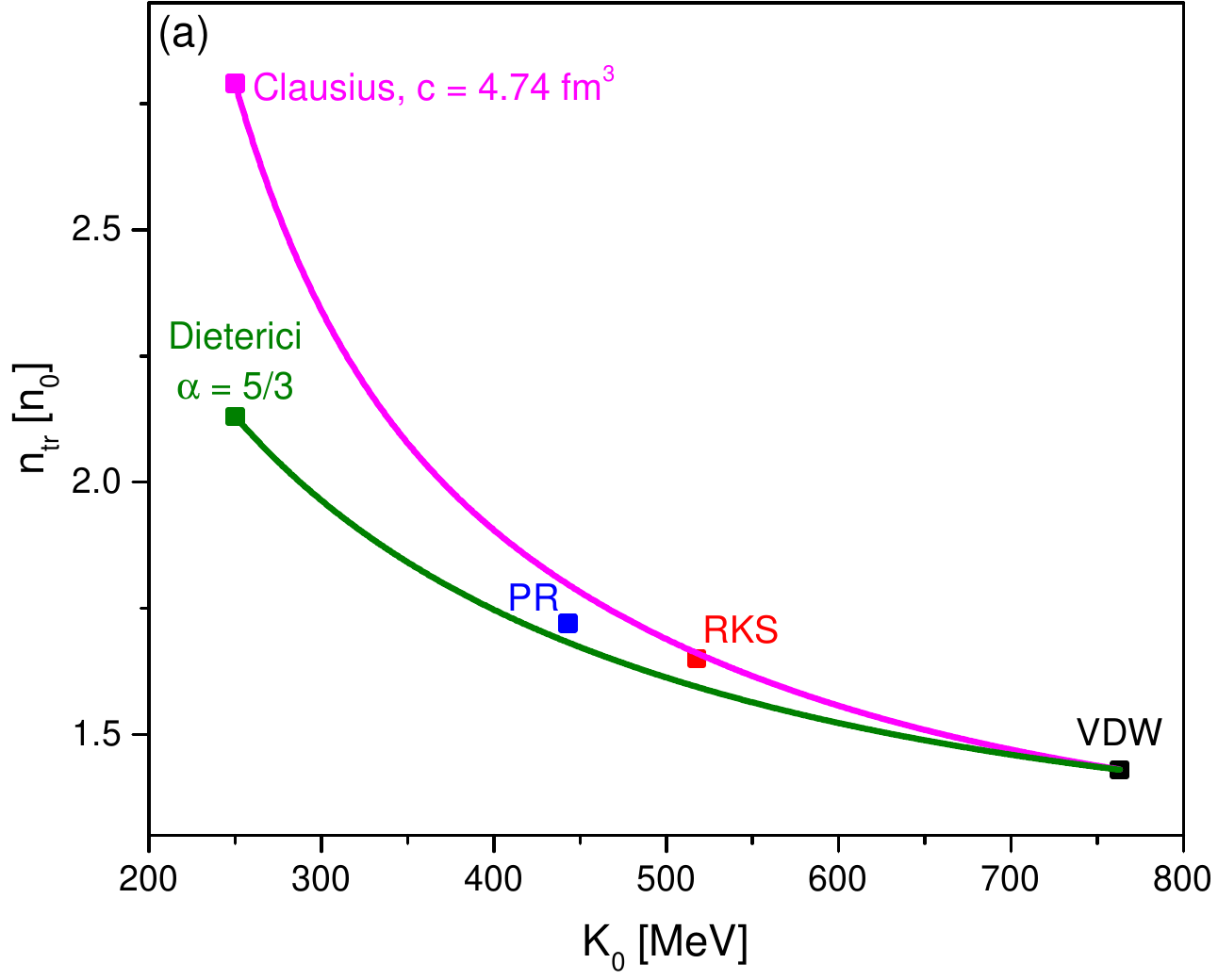}
    \includegraphics[width=.49\textwidth]{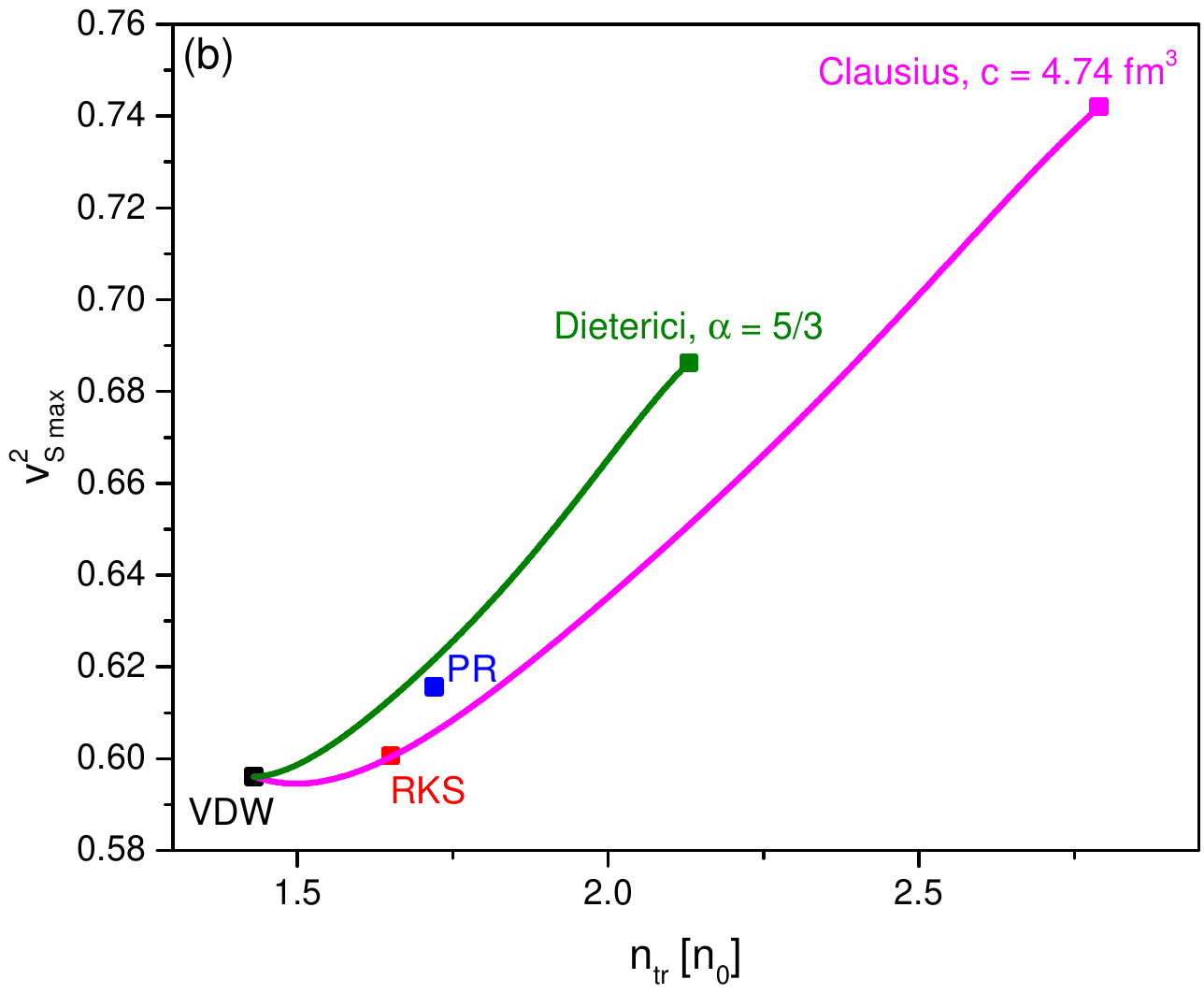}
    \caption{ (a) The transition density to quarkyonic regime $n_{tr}$ as the function of incompressibility $K_0$ and (b) the value of the speed of sound at the peak $v^2_{s,{\rm max}}$ as the function of $n_{tr}$ in the vdW, RKS, PR, Clausius, and Dieterici models.}
    \label{fig45}
\end{figure*}

Here quarks are viewed as non-interacting, $n_Q=n_Q^{\rm id}$, $\varepsilon_Q=\varepsilon_Q^{\rm id}$, and at $T=0$ a sharp phase separation in momentum space is implemented with quarks occupying low momentum states from zero momentum $q=0$ to quark Fermi momentum $q_{\rm bu}$. Thus, quark contributions to baryon and energy densities read
\eq{\label{nq}
n_Q & = \frac{g}{2\pi^2} \int_0^{k_{\rm bu}/N_c} q \sqrt{q^2 + \Lambda^2} dq \ , \\\label{eq}
\varepsilon_Q & = \frac{g N_c}{2\pi^2} \int_0^{k_{\rm bu}/N_c} q\sqrt{q^2 + (m/N_c)^2} \sqrt{q^2 + \Lambda^2} dq.
}

Here $k_{\rm bu}/N_c=q_{\rm bu}$ is a quark Fermi surface, $N_c=3$ is the color degeneracy factor, $m/N_c$ is the quark mass, $g=4$ is the spin-isospin degeneracy factor,\footnote{The quarks include an extra degeneracy factor of $N_c$ from color. While this factor cancels out with the $1/N_c$ fractional baryon number in the expression~(\ref{nq}), however remains in the energy density~(\ref{eq}).} and $\Lambda$ is the infrared regulator which smooths out the sudden onset of quarks. This is necessary to prevent an acausal speed of sound, and here we take $\Lambda=306~$MeV in accordance with Ref.~\cite{Pang:2023dqj}.

Nucleons occupy a shell surrounding the quark Fermi sea, with momenta lying between $k_{\rm bu} < k < k_F$. Thus, for noninteracting nucleons one has
\eq{
n_{N}^{\rm id} & = \frac{g}{2\pi^2}  \int_{k_{\rm bu}}^{k_F} k^2 dk = \frac{g}{6\pi^2} [(k_F)^3 - (k_{\rm bu})^3] \ ,\\
\varepsilon_N^{\rm id} & = \frac{g}{2\pi^2}  \int_{k_{\rm bu}}^{k_F} k^2\sqrt{k^2 + m^2} dk \ .
}

The interactions between nucleons are taken into account by Eqs.~(\ref{P-Tmu})--(\ref{e-Tmu-1}). In particular, for Dieterici and Clausius models the total energy density reads
\eq{
\varepsilon_{\rm D} & = \varepsilon_Q + (1-bn_{N})\varepsilon_{N}^{\rm id} - a\frac{n_{N}^{\alpha}}{\alpha - 1},
\\
\varepsilon_{\rm C} & = \varepsilon_Q + (1-bn_{N})\varepsilon_{N}^{\rm id} - \frac{a n_{N}^2}{1+c n_{N}}.
}

At a given baryon density, the values of $k_{\rm bu}$ and $k_F$ are determined by minimizing the energy density of the system. 

Finally, the speed of sound is calculated as
\eq{
v_s^2=\frac{n}{\mu_B}\frac{d \mu_B}{d n}=\frac{n}{\mu_B}\frac{d^2 \varepsilon}{d n^2}~.
}

During the hadronic phase, repulsive interactions lead to a rapidly stiffening equation of state as the density of the system increases. This corresponds to a sharp rise in the speed of sound which surpasses the conformal limit. The appearance of free quarks softens the equation of state, leading to a peak in the speed of sound which then falls back below the conformal limit. For this reason, we use the peak in the speed of sound as an indicator of a transition to the quarkyonic regime and define the quarkyonic transition density $n_{tr}$ as the density at which the peak in the speed of sound occurs, $v_s^2|_{n_{tr}}=v_{s,{\rm max}}^2$.

Figure~\ref{figv2snn0} shows $v_s^2$ as a function of scaled baryon density for each parametrization of attraction. The Dieterici and Clausius formulations are both taken at their limiting values for our range of investigation, $\alpha=5/3$ and $c=4.74~ \rm fm^3$, respectively. One sees that for all considered models $v_s^2$ at the peak significantly exceeds the conformal limit of $1/3$; however, it remains within the constraint of causality, $v^2_{s,{\rm max}}<1$.

One also observes the anticorrelation between $K_0$ and $n_{tr}$. Namely, the models with lower $K_0$ exhibit a peak in $v^2_{s,{\rm max}}$ which occurs at higher densities and has a broader shape. This corresponds to a later and more gradual appearance of quarks. We present this correlation between $K_0$ and $n_{tr}$ in Fig.~\ref{fig45}(a). One sees that for empirical range of $K_0 \simeq 250 - 315$~MeV, the peak in $v_s^2$ occurs at densities of $n_{tr} \approx$ (1.9-2.1)$n_0$ and $n_{tr} \approx$ (2.3-2.8)$n_0$ for Dieterici and Clausius formulations, respectively. Each of these smoothly approach the van der Waals case as $\alpha\rightarrow2$ and $c\rightarrow0$, with transition densities for the Clausius model trending above that of Dieterici as they converge to the van der Waals limit.

The anticorrelation between $n_{tr}$ and $K_0$, which is depicted in Fig.~\ref{fig45}(a), can be well approximated by the formula $n_{tr} = AK_0^{-3/2} + B$. 
In the case of the Clausius model (\ref{P_CTn}), the approximation gives $A = (6507.191 \pm 100.145) \ \rm MeV^{3/2}$, $B = (1.105 \pm 0.015) \ n_0$, and in the case of the Dieterici model (\ref{P_DTn}) one has $A = (3412.042 \pm 118.335) \ \rm MeV^{3/2}$, $B = (1.295 \pm 0.017) \ n_0$.

From Fig.~\ref{figv2snn0} one also sees the anticorrelation between incompressibility $K_0$ and the height of the peak in the speed of sound $v^2_{s,{\rm max}}$: the lower values of $K_0$ correspond to higher $v^2_{s,{\rm max}}$. As both $n_{tr}$ and $v^2_{s,{\rm max}}$ are anticorrelated with $K_0$ they are correlated with each other. This correlation is presented in Fig.~\ref{fig45}(b), where $v^2_{s,{\rm max}}$ is shown as a function of $n_{tr}$.

\section{Summary}\label{sum}

We considered nuclear matter within different real gas models (\ref{P_VDW})--(\ref{P_DTn}) and wrote down the corresponding equations of state in the grand canonical ensemble with the introduction of quantum statistics (\ref{P-Tmu})--(\ref{p-id}). By requiring each EoS to satisfy the nuclear matter ground-state conditions (\ref{n0}) and (\ref{W0}), we have shown that there is a correlation between the incompressibility $K_0$, the critical temperature $T_c$, and the critical particle number density $n_c$ (see Figs. \ref{fig2} and \ref{fig3}). In addition, we also applied the above equations of state to describe the nucleonic part in the quarkyonic matter model. We found that the dynamically generated transition to quarkyonic matter occurs at a density $n_{tr}$ which is anticorrelated with nuclear incompressibility $K_0$.

We emphasize that the parameters $a$ and $b$ responsible for attraction and repulsion, respectively, for all equations of state considered are uniquely fixed by fitting the model to the ground state of nuclear matter (\ref{n0}) and (\ref{W0}). Thus, for any equation of state we have the same values of thermodynamic functions at $T = 0$ and $n_0 = 0.16 \ \rm fm^{-3}$ by construction. This fact brings a certain self-consistency to the mutual simultaneous change of parameters $a$ and $b$, when varying the so-called third parameters $\alpha$ or $c$ in models (\ref{P_CTn}) and (\ref{P_DTn}): changing one of the parameters ($a$ or $b$), the other must change strictly in accordance with (\ref{n0}) and (\ref{W0}), so as not to disturb the ground state of nuclear matter.

Such a consistent change in the magnitudes of attraction and repulsion interactions is manifested in the correlation between the incompressibility factor and the position of the CP (see Figs. \ref{fig2} and \ref{fig3}): an increase in the incompressibility factor $K_0$ leads to an increase in the critical temperature $T_c$ and the critical density of the number of particles $n_c$. We emphasize that there is not only a $T_c \leftrightarrow K_0$ correlation, but also $n_c \leftrightarrow K_0$ and, as a consequence, $n_c \leftrightarrow T_c$.

The nuclear incompressibility also strongly correlates to the onset of quarkyonic matter. At $T=0$, this transition is driven by energy minimization; at large densities, free quark states become energetically favorable due to hadronic interactions. At low values of incompressibiliy, i.e., $c=4.74 ~ \rm fm^3$ and $\alpha=5/3$, the EoS stiffens more gradually with increasing density during the hadronic phase. This leads to a later appearance of free quarks, a higher peak in the speed of sound, and a broader width of this peak. As we lower the value of $K_0$ and converge towards the vdW case, the transition density to quarkyonic matter decreases monotonically, as does the height of the peak in $v_s^2$ (see Figs.~\ref{figv2snn0}~and~\ref{fig45}). Despite these quantitative differences among the various models investigated, we still see the same characteristic signatures of a transition to quarkyonic matter, namely a rapid stiffening of the EoS during the hadronic phase, causing a sharp rise in the speed of sound exceeding the conformal limit. As an outlook, it may be interesting to explore the properties of quarkyonic matter at nonzero isospin asymmetries. Such an analysis was done in Ref.~\cite{Moss:2024uam} for the vdW equation, and it would be interesting to explore how the lower $K_0$ values in Dieterici or Clausius models may affect the resulting properties of neutron-star matter. \\

{\bf Acknowledgements.}
V.V. thanks Agnieszka Sorensen for valuable remarks.
M.I.G. is thankful for support from the Simons  Foundation.

\appendix

\section{Critical point location in the Boltzmann approximation} \label{appA}

In the classical van der Waals model (\ref{P_VDW}), Eq.~(\ref{CP1}) gives the well-known expressions (see, e.g., Ref.~\cite{Landau:1980mil})
\eq{ \label{VDW-ab}
T_c^0 = \frac{8a}{27b}~,~~~~
 n_c^0 = \frac{1}{3b}~.
}
Straightforward calculations of Eq. (\ref{CP1}) lead to the following analytical expressions for $T_c^0$ and $n_c^0$: \\
in the classical Redlich-Kwong-Soave model (\ref{P_RKSTn})
\eq{ \label{RKS-ab}
 T_c^0 = \frac{3(\sqrt[3]{2} - 1)^2a}{b}~,~~~
 n_c^0 = \frac{\sqrt[3]{2} - 1}{b}~,
}
in the classical Peng-Robinson model (\ref{P_PRTn})
\eq{\label{PR-ab}
 & T_c^0 = d_1\frac{a}{b}~,~~~
 n_c^0 = d_2\frac{1}{3b}~,\\
 & d_1 \approx 0.17~, ~~~d_2 \approx 0.76~,
}
in the classical Clausius model (\ref{P_CTn})
\eq{\label{C-ab}
T_c^0 = \frac{8a}{27(b + c)}~,~~~
& n_c^0 = \frac{1}{3b + 2c}~,
}
and in the classical Dieterici model (\ref{P_DTn})
\eq{\label{D-ab}
& T_c^0 = \frac{4a \alpha}{(\alpha + 1)^2}\left( \frac{1}{b} \frac{\alpha - 1}{\alpha + 1} \right)^{\alpha - 1}~,\\
& n_c^0 = \frac{1}{b} \frac{\alpha - 1}{\alpha + 1}~.
}

The $T_c$ and $n_c$ values with Fermi statistics for the above models are calculated numerically and presented in Tables \ref{table-1}--\ref{table-2-2} and Figs.~\ref{fig2} and \ref{fig3}. 

\bibliography{references}

\end{document}